\documentclass{eas}
\usepackage{graphicx}
\usepackage{sidecap}
 \newcommand{\apj}         {ApJ}
 
 \newcommand{\aap}         {A\&A}
 \newcommand{\araa}        {ARAA}
 \newcommand{\mnras}       {MNRAS}
 \newcommand{\nat}         {Nature}
\newcommand{\hi}          {\mbox{\rm H{\small I}}}

\newcommand{\htwo}        {\mbox{H$_{2}$}}
\newcommand{\jone}        {\mbox{$J=1\rightarrow0$}}

\newcommand{\msunperpcsq} {\mbox{\rm M$_\odot$~pc$^{-2}$}}

\newcommand{\xco}         {\mbox{$X_{\rm CO}$}}

\newcommand{\aco}         {\mbox{$\alpha_{\rm CO}$}}
\newcommand{\xcounits}    {\mbox{\rm cm$^{-2}$(K km s$^{-1}$)$^{-1}$}}
\newcommand{\acounits}  {\mbox{\rm M$_\odot$ (K km s$^{-1}$ pc$^2$)$^{-1}$}}

\newcommand{\av}          {\mbox{$A_V$}}

\begin{document}

\title{Using CO to Measure Molecular Masses} 
\author{Alberto D. Bolatto}\address{Department of Astronomy, University of Maryland, College Park, MD 20742, USA}
%
%
\begin{abstract}
With an increased appreciation for the role of gas in galaxy
evolution, there is renewed interest in measuring gas masses for
galaxies. I review some of the basic concepts in using CO to determine
molecular masses, and discuss some of the recent work.
\end{abstract}
\maketitle
\section{Introduction}
Observations of rotational CO transitions are the primary method used to
determine molecular masses, particularly in systems outside the Milky
Way. For an expanded discussion of using CO to determine masses, I
refer the reader to the review by Bolatto, Wolfire, \& Leroy
(\cite{BOLATTO2013}). Here I briefly introduce some of the concepts
and focus on work published since the
review.

The relationship between molecular column density or mass and $^{12}$CO
\jone\ emission is determined through two equations:

\begin{eqnarray}
  {\rm N(H_2)\,[cm^{-2}]} &=& {\rm X_{CO}\,I_{CO}\,[K\,km\,s^{-1}]} \\
  {\rm M_{mol}\,[M_\odot]} &=& {\rm \alpha_{CO}\,L_{CO}\,[K\,km\,s^{-1}\,pc^2].}
\end{eqnarray}

\noindent In the disk of the Milky Way the value of
\xco\ is $\approx2\times10^{20}$~\xcounits, which in turn implies
$\aco=4.36$~\acounits\ including a contribution to the mass of 36\%
due mostly to He according to its cosmological abundance.

\section{Physics of the CO-to-H$_2$ Conversion Factor} 
The $^{12}$CO \jone\ transition is usually very optically thick (in
the Milky Way disk the typical $^{12}$CO/$^{13}$CO ratio is
$\sim10-15$, suggesting $\tau_{\rm CO}>5$). For gravitationally-bound
clouds, its integrated intensity encodes information about mass
through the line width, so the use of these equations requires that CO
and \htwo\ are coextensive and that the velocity dispersion of the gas
is a reflection of its self-gravity. The peak temperature of the line
carries information about the product of the temperature of the gas
and its beam-filling fraction. It is easy to show that for
self-gravitating entities \xco\ will have mild dependencies on density
and temperature, such that $\xco\sim n^{0.5}\,T_K^{-1}$ where $n$ is
the gas density and $T_K$ is its kinetic temperature (see Bolatto et
al. \cite{BOLATTO2013} for a more detailed discussion). Note that
since $\tau_{\rm CO}\gg1$, the relevant $T_K$ is that at the
$\tau_{\rm CO}=1$ surface.

The fact that CO \jone\ is optically thick makes this method robust to
changes in the CO abundance and to the details of the excitation
(\S\ref{sec:breakdown}), which affect methodologies that rely on
optically thin tracers. By comparison, in the optically thin regime
$\xco\sim[\htwo/{\rm CO}]\,T_{ex}\,\exp(h\nu/(kT_{ex}))$ (a similar
equation can be written for any optically thin rotation
transition). Determining masses using optically thin emission requires
accurate knowledge of abundances and excitation temperatures, either
of which may change over a very large range (for example, $T_{ex}$ is
a very strong function of density and $T_K$).

Using CO to estimate molecular masses of entire galaxies requires
further assumptions (Dickman, Snell, \& Schloerb \cite{DICKMAN1986}):
the CO emission must arise from an ensemble of self-gravitating clouds
with a narrow range of $n^{0.5}\,T_K^{-1}$. As long as the velocity
dispersion in each cloud in the ensemble is a reflection of its
gravity, and clouds do not shadow each other, it can be shown that the
total luminosity will reflect the sum of the cloud masses.

\section{Break Down of the Standard Conversion}
\label{sec:breakdown}
The assumptions behind the standard proportionality between
CO luminosity and molecular mass break down in three
circumnstances: 1) at low metallicities, 2) in environments where the
gas motions do not reflect self-gravity, and 3) in places where
$n^{0.5}\,T_K^{-1}$ is very different from the disk of the Milky Way.

Although (3) may be important in some cases, it is probably not the
dominant driver of changes. Because $T_K$ depends on the radiation
field --- mostly driven by star formation --- and star formation is
enhanced at high density, there is likely a large degree of
compensation in their ratio. Moreover, photodissociation regions
(PDRs) are self-regulated to keep an approximately constant
temperature for the $\tau_{\rm CO}=1$ surface; increasing the
radiation field impinging on a PDR drives its surface temperature
higher, but it also pushes the ${\rm C^+}\rightarrow{\rm CO}$
transition further into the cloud and into cooler regions.

Realistic simulations incorporating cloud structure and
chemistry provide key insights on how \xco\ is affected by changes in
the radiation field and the distribution of column densities, velocity
dispersion, and dust extinction (e.g., Clark \& Glover
\cite{CLARK2015}). Cloud structure has an impact on the effects of
increasing the incident radiation field, but \xco\ changes little in
large, dense, and bound clouds.

\subsection{Low Metallicities}
\label{sec:lowmet}
At low metallicities the hypothesis of coextensivity between H$_2$ gas
and CO emission breaks down. High CO abundance requires a minimum
extinction ($\av\sim1-2$, weakly dependent on the incident
radiation field). By comparison, most of the atomic hydrogen becomes
molecular at a much lower extinction ($\av\sim0.2$). Extinction is
more difficult to build up at low metallicity, where the dust-to-gas
ratio is lower, resulting in a increasing fraction of the H$_2$ gas
not being associated with bright CO emission.
Physical models that incorporate the relevant physics qualitatively
agree on this picture (Maloney \& Black \cite{MALONEY1988};
R\"ollig et al. \cite{ROLLIG2006}; Wolfire, Hollenbach, \& McKee
\cite{WOLFIRE2010}), as do numerical calculations incorporating cloud
structure and time-dependent chemistry (Glover \& Mac Low
\cite{GLOVER2011}).

Resolved observations of the Magellanic Clouds further validate this
picture; a strong correlation between CO intensity and inferred
\av\ is seen (Lee et al. \cite{LEE2015}), and global \aco\ values
rapidly increase for decreasing metallicity (e.g., Jameson et
al. \cite{JAMESON2015}). Indeed, the CO emitting regions of clouds at
low metallicity are increasingly smaller for decreasing metallicity,
while the fraction of the cloud encompassed by the molecular envelope
faint in CO increases (Leroy et al. \cite{LEROY2007},
\cite{LEROY2009}). This is strikingly demonstrated in the recent
observations of individual molecular clouds in the very low
metallicity WLM galaxy, which show very small CO emitting regions
(Rubio et al. \cite{RUBIO2015}).

Thus at low metallicity the amount of CO emission per \htwo\ molecule
is set by the distribution of column densities in the cloud, mediated
by the dust-to-gas ratio. It is, in fact, fairly insensitive to the
gas metallicity itself, but it depends on metallicity through the
dust-to-gas ratio. This highlights the importance of better
understanding dust production and destruction mechanisms in
galaxies. It is increasingly clear that there is not a one-to-one
correspondence between gas metallicity and dust-to-gas ratio (e.g., Fisher et
al. \cite{FISHER2014}), with observations finding a large dispersion
in dust-to-gas ratio at fixed metallicity for low metallicity and high
specific star formation rate (Remy-Ruyer et
al. \cite{REMY-RUYER2014}). Understanding how the distribution of
cloud column densities is established is also important, since it
plays a key role at determining the fraction of the gas that reaches
the critical \av\ necessary for bright CO emission.

\subsection{Increased line-widths not due to self-gravity}

\begin{figure}[t]
   \centering
   \resizebox{9cm}{!}{
 	 {\includegraphics{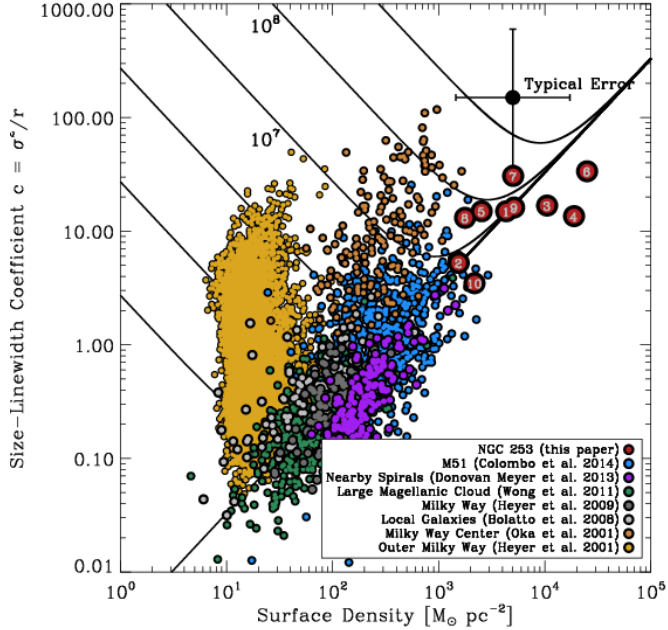}}}
   \caption{ Figure 13 from Leroy et al. (\cite{LEROY2015}) showing
     the physical state of a large sample of Giant Molecular Clouds
     (GMCs) in galaxies, including clouds identified in the starburst
     of NGC~253 (numbered circles). The curves illustrate the
     theoretical lines for clouds in equilibrium with external
     pressure (Field, Blackman, \& Keto \cite{FIELD2011}). The
     diagonal is, essentially, the classical virial equilibrium. Using
     a combination of optically thin tracers the GMCs in NGC~253 are
     placed in this diagram. The resulting value of \aco\
     for these structures is somewhat uncertain but
     close to Galactic, with a median value of $\aco\sim3$~\acounits.\label{fig:one}}
 \end{figure}

In very gas-rich galaxies it is possible to maintain an extended
intercloud medium that is molecular and emits in CO. This medium experiences the combined gravitational potential of the stellar,
gas, and dark matter components, not just its own self-gravity. In
these circumnstances the hypothesis that the velocity dispersion of
the medium is a reflection of its mass is broken, and a ``diffuse''
intercloud medium may dominate the CO
luminosity. A simple-minded application of the CO-to-\htwo\ conversion
factor yields the harmonic mean between the molecular and stellar
masses. This is likely the reason for the well-established observation
that local ultra-luminous IR galaxies are disproportionally luminous
in CO (Downes \& Solomon \cite{DOWNES1998}), although a dense phase
could still dominate their gas mass (Papadopoulos et
al. \cite{PAPADOPOULOS2012}).

A particularly interesting case to study is the NGC~253 starburst,
close enough to observe in great detail with the Atacama Large
Millimeter Telescope (ALMA). Indeed, ALMA observations can
simultaneously be used to measure the global properties and also break
up the emission into clouds. Leroy et al. (\cite{LEROY2015}) present
observations of the nuclear region where high density tracer molecules
(HCN, HCO$^+$, CS, and isotopologues) are used together to identify
ten giant molecular clouds in the starburst.  The authors then use six
optically thin tracers (which include both molecules and dust) to
determine masses for each cloud: the different tracers have a factor
of 3 scatter, but a well defined mean. Using the surface densities so
determined together with the size and velocity dispersion measurements,
they are able to place these clouds in a virial diagram (Figure
\ref{fig:one}).

The result is that the clouds in the NGC~253 starburst are
self-gravitating, in the sense that their velocity dispersion reflects
the expectation for virialized clouds of their surface density. Cloud
surface densities are very large, hovering around $10^4$
\msunperpcsq\ (clouds in the Milky Way disk are
typically $\sim10^2$ \msunperpcsq). Nonetheless, the derived
conversion factor is $\aco\sim3$~\acounits. The conversion factor in
the large clouds is even higher, with the four largest clouds yielding
$\aco\sim4.5$~\acounits. The uncertainties are large (0.5 dex), due to
the systematic uncertainties in the optically thin mass determination,
and the na\"{\i}ve application of the $n^{0.5}\,T_K^{-1}$ scaling
would lead us to expect a factor $2-3$ higher conversion factor. These
clouds are also embedded in an extended CO ``diffuse'' medium, which
has most of the luminosity. The total measured \aco\ is uncertain but
likely $\aco\sim 1$~\msunperpcsq.

This same picture of a mix of phases probably accounts for the bulk of
the discrepancy between \aco\ estimated using the virial theorem in
resolved clouds in some galaxy centers (Donovan-Meyer et al.
\cite{DONOVANMEYER2013}) and \aco\ estimated from dust modeling in
the same objects (Sandstrom et al. \cite{SANDSTROM2013}). In summary,
what matters in these circumnstances is to understand what fraction of
the total CO emission originates from self-gravitating complexes, and
what is the source of the observed velocity dispersion. Bound,
self-gravitating complexes will likely have a conversion factor that
is close to the ``Galactic'' value (modulated by changes due to
density and temperature effects), while the intercloud medium could
contribute significantly to the luminosity.

\section{Conclusions}
In Bolatto et al. (\cite{BOLATTO2013}) we propose the following
expression to encompass the regimes of the conversion factor:

\begin{equation}
\aco \approx 2.9\exp \left(\frac{+0.4}{Z^\prime~\Sigma_{\rm GMC}^{100}} \right) 
\left(\Sigma_{\rm total}^{100}\right)^{-\gamma}\,,
\label{eq:aco}
\end{equation}

\noindent with $\gamma\approx0.5$ for $\Sigma_{\rm
  total}>100$~\msunperpcsq\ and $\gamma=0$ otherwise. The gas
metallicity relative to the Milky Way, $Z'$, is used here as an
``observable'' proxy of the dust-to-gas ratio, but we have pointed out
the shortcomings of that approximation in \S\ref{sec:lowmet}. The
``typical'' surface density of clouds in the Milky Way disk is assumed
to be 100~\msunperpcsq, hence $\Sigma^{100}_{\rm GMC}$ is the typical
surface density of molecular clouds in the system relative to that in
the Milky Way disk. Conversely, $\Sigma_{\rm total}^{100}$ stands for the
total surface density that is exerting gravitational attraction in units of
100~\msunperpcsq.

The first factor in Equation \ref{eq:aco} incorporates the physics of
the \hi-\htwo\ transition relative to the C$^+$-CO transition, as we
understand them. In systems of lower metallicity \aco\ will increase
exponentially unless the decrease in metallicity is compensated by an
increase in the typical surface density of clouds. The exponential
character of the relation is not a fitting choice: it reflects the
theoretical expectation that CO emission is rapidly confined to the
highest column density regions of the cloud (Wolfire et
al. \cite{WOLFIRE2010}). In that sense, CO acts more as a column
density tracer than as a bulk mass tracer at low metallicity.

The second factor in Equation \ref{eq:aco} is considerably more
uncertain, and reflects the expectation that the CO emitting gas,
while not necessarily self-gravitating, still has to be bound to the
overall potential of the system. Therefore its velocity dispersion
will be a function of the total surface density.  The crucial unknown,
which we repeat, is the fraction of CO emission that arises from
self-gravitating clouds: a system where most CO emission originates in
such clouds will likely have \aco\ close to Galactic.  The choice of
100~\msunperpcsq\ as the break point between purely self-gravitating
gas and emission from gas bound to the overall potential is likely
conservative (it could be higher), and put forward based on the
results for galaxy centers.  Similarly, the $\gamma=0.5$ exponent
implicitly assumes that the CO luminosity is dominated by the
``diffuse'' intercloud phase. Galaxy centers may be peculiar: they can
have high apparent velocity dispersions due to bulk gas motions
(streaming due to bars, for example) which would result in a
disproportionately large fraction of ``diffuse'' CO emission. So it is
possible that this second factor incorrectly predicts too large a
correction in \aco. Conversely, it would be hard to push \aco\ under
the value of this prediction unless there is a large increase in the
gas temperature. Undoubtedly, future observations of optically thin
tracers and multi-transition excitation studies will help clarify some
of these unknowns.


\end{document}